\documentclass[a4paper, 10pt, dvips]{article}

\usepackage{graphicx}
\usepackage{amssymb}
\usepackage{amsmath}
\newcommand {\bR} {\mathbb{R}}

\newcommand {\fexp} [1] {\exp \left( #1 \right)}

\newcommand {\fabs}[1] {\left| #1 \right|}
\newcommand {\fabsq}[1] {\left| #1 \right|^2}
\newcommand {\fsqrt}[1] {\sqrt{#1} \,}

\newcommand {\fnormq} [1] {{\left\| #1 \right\|}^2}

\newcommand{\intRRR}{\int_{\mathbb{R}^3} d\vec{x} \,}

\newcommand{\refeq}[1]{eq. (\ref{#1})}

\newcommand {\Ang}{\mbox{\AA}}

\begin{document}
\title{Relativistic Time of Arrival} 
\author{Andreas Ruschhaupt\thanks{Email:
    rushha@physik.uni-bielefeld.de} \\ \\ {\small
    Faculty of Physics, University of Bielefeld,}\\ {\small
    Universit\"atsstr. 25, D-33615 Bielefeld, Germany}} 
\date{} 
\maketitle

\begin{abstract}
  We propose a covariant algorithm for relativistic ideal measurements and
  for relativistic continuous measurements, its non-relativistic limit results
  the algorithm of the Event-Enhanced Quantum Theory. Therefore an additional
  intrinsic parameter, the proper time, is used.
  As an application we compute the time of arrival of a particle at
  a detector and find good agreement between the expected values of the time
  of arrival for weak detectors and the results of the relativistic
  point-mechanic over a wide range.
  For very high momentums there is a small probability for a negative
  time of arrival, so the expected times are a bit smaller than the results of
  the relativistic mechanics. 
\end{abstract}

%
%

\section{Introduction}

One can use the PDP-algorithm of the Event-Enhanced Quantum Theory (EEQT) as
described by Blanchard and Jadczyk \cite{blanchard.1995a, blanchard.1995b,
  blanchard.1995c, blanchard.1995d} to simulate detections of a non-relativistic
electron (for example \cite{blanchard.1996a, blanchard.1998,
  ruschhaupt.1998}).

In this paper we are interested in detections of a relativistic electron in an
external electromagnetic field.

Trying to define states and a reduction postulate in a relativistic theory can imply
a lot of paradoxes and difficulties (for example see Y. Aharonov and D.Z. Albert
\cite{aharonov.1980, aharonov.1981, aharonov.1984}).

One possibility to avoid (some) difficulties is to consider the wave function
for relativistic particle not as a function on the space-time continuum but as a
function on the set of flat, space-like hypersurfaces in Minkowski space (for
example see the papers by Breuer and Petruccione \cite{breuer.1998a, breuer.1998b}).

Another possibility is the introduction of a supplementary,
intrinsic time, the proper time (for example see the paper by Horwitz and Piron
\cite{horwitz.1973} or the review paper by Fanchi \cite{fanchi.1993}).

Blanchard and Jadczyk \cite{blanchard.1996b} formulated a relativistic algorithm
for events by using this proper time.

In the following we also use the proper time to propose a relativistic extension of
the EEQT, but another definition for the state of the system and its
dynamics.

The total system consists of a classical and a quantum part. At a given proper
time $\tau$
the state of the total system is a pair $(\omega_\tau,\Psi_\tau)$, $\omega_\tau$
is the state of the classical and $\Psi_\tau$ is the state of the quantum part.

We assume, that the classical part has only a finite number $N_C$ of possible
pure states, therefore a state $\omega_\tau$ of the classical part is a number
$\omega_\tau \in \{0..N_C-1\}$.

A state $\Psi_\tau$ of the quantum part must have the following properties:

\begin{list}{}{\topsep0cm}
\item [(i)] $\Psi_\tau$ is continuously differentiable and a solution of the Dirac
  equation with an external electromagnetic field
\begin{eqnarray}
  \left(\imath \gamma^\mu \partial_\mu - \frac{e}{c\hbar} \gamma^\mu A_\mu -
    \frac{mc}{\hbar}\right) \Psi_\tau = 0
\end{eqnarray}

\item [(ii)] For all $y = (y^0, \vec{y}) \in \bR^4$, $\vec{\alpha} \in \bR^3$
  with $\fabsq{\vec{\alpha}} < 1$ and $\vec{\varphi} \in \bR^3$ with
  $\fabs{\vec{\varphi}} < 2\pi$ it follows:
\begin{eqnarray} 
  \intRRR \fabsq{\Psi_\tau (y^0 + \vec{\alpha} \cdot (R(\vec{\varphi})\vec{x}),
    \vec{y} + R(\vec{\varphi})\vec{x})} < \infty \\ \lim_{\fabs{\vec{x}} \to
    \infty} \fabs{\vec{x}} \fabsq{\Psi_\tau (y^0 + \vec{\alpha} \cdot
    (R(\vec{\varphi})\vec{x}), \vec{y} + R(\vec{\varphi})\vec{x})} = 0
\end{eqnarray}
\end{list}
In this paper we use the Dirac representation of the $\gamma$-matrices.
Moreover $R(\vec{\varphi}) \in SO(3)$ should be the rotation of
$\fabs{\vec{\varphi}}$ around the vector $\vec{\varphi}/\fabs{\vec{\varphi}}$.

A quantum state is uniquely given by its values on a plane
\begin{eqnarray}
  P_{(y^0,\vec{y}),\vec{\alpha},\vec{\varphi}} = \{(y^0 + \vec{\alpha} \cdot
  (R(\vec{\varphi})\vec{x}), \vec{y} + R(\vec{\varphi}) \vec{x}) \, | \, \vec{x}
  \in \bR^4\}
\end{eqnarray}
with $y = (y^0, \vec{y}) \in \bR^4$, $\vec{\alpha} \in \bR^3,
\fabsq{\vec{\alpha}} < 1$ and $\vec{\varphi} \in \bR^3$ with
$\fabs{\vec{\varphi}} < 2\pi$.

We introduce the operator $U^{-1}((y^0, \vec{y}), \vec{\alpha},\vec{\varphi})$, which reduces the
quantum state to a ''plane state'':
\begin{eqnarray}
  (U^{-1} ((y^0, \vec{y}), \vec{\alpha}, \vec{\varphi}) \Psi) (\vec{x}) :=
  \Psi (y^0 + \vec{\alpha} \cdot (R(\vec{\varphi})\vec{x}), \vec{y} + R(\vec{\varphi}) \vec{x})
\end{eqnarray}

The operator $U^{-1}((y^0, \vec{y}), \vec{\alpha},\vec{\varphi})$ is invertible,
we call the inverse
operator $U((y^0, \vec{y}), \vec{\alpha},\vec{\varphi})$. Both operators are unitary.

In the following, we examine how the state of the total system changes, if we
change the reference frame $K \to \tilde{K}$ with $\tilde{x} = \Lambda x + a$.
We look only at Poincar\'e-transformations $(\Lambda, a)$ which do not mirror the space or
invert the direction of time.

The classical state and the proper time should be invariant. The quantum state
changes in the following way
\begin{eqnarray}
  \Psi \longrightarrow \tilde{\Psi}(\tilde{x}) = S(\Lambda) \Psi (\Lambda^{-1}
  (\tilde{x} - a))
\end{eqnarray}
$S$ being a non-singular $4 \times 4$ matrix with $S(\Lambda) \gamma^\mu
S^{-1}(\Lambda) = (\Lambda^{-1})^\mu_{\;\;\nu} \gamma^\nu$.

Now we introduce a scalarproduct between two quantum states by
\begin{eqnarray}
  \lefteqn{<\Psi_A|\Psi_B>} & & \nonumber \\ & = & \int_{\mathbb{R}^3} d\vec{x} \;
  \Psi_A^+ (y^0 + \vec{\alpha} \cdot (R(\vec{\varphi})\vec{x}), \vec{y} +
  R(\vec{\varphi})\vec{x}) \left[ 1 - \gamma^0 \vec{\gamma} \vec{\alpha} \right]
  \Psi_B (y^0 + \vec{\alpha} \cdot (R(\vec{\varphi})\vec{x}), \vec{y} +
  R(\vec{\varphi})\vec{x})
\end{eqnarray}
with $y=(y^0,\vec{y}) \in \mathbb{R}^4$, $\vec{\alpha} \in \mathbb{R}^3$ with
$\fabsq{\vec{\alpha}} < 1$ and $\vec{\varphi} \in \bR^3$ with
$\fabs{\vec{\varphi}} < 2\pi$ arbitrary.

The scalarproduct is positive definite and well defined, it is independent of
the choose of $y^0, \vec{y}, \vec{\alpha}, \vec{\varphi}$. Note that the number
of free parameters (10) equals the number of parameters of a
Poincar\'e-transformation.

Moreover it is covariant, its value is equal in all reference frames
$<\Psi_A|\Psi_B>_K = <\tilde{\Psi_A}|\tilde{\Psi_B}>_{\tilde{K}}$.

Events like the preparation or the detection of an electron happen at a proper
time $\tau_i$ at a space-time point $x_i$.

To preserve a kind of order, we assume the following: taking two events happen at
$\tau_1$ and $\tau_2$ with $\tau_1 < \tau_2$, there must be a reference frame,
in which the time of the first event $\tilde{x_1}^0$ is earlier than the time of
the second event $\tilde{x_2}^0$ (we only allow Poincar\'e-transformations,
which do not mirror the space or invert the direction of time).

Therefore no ''later'' event can take place in the backward light-cone of a
previous event:
\begin{eqnarray}
  \tau_1 < \tau_2 \Rightarrow ((\fnormq{x_2 - x_1} \ge 0 \; \mbox{and} \; x_1^0 <  x_2^0)
  \; \mbox{or} \; (\fnormq{x_2 - x_1} < 0)
  \label{eq_order}
\end{eqnarray}
$\fnormq{x} = \fnormq{(x^0,\vec{x})} = (x^0)^2 - \fabsq{\vec{x}}$ being the
Minkowski-distance.

This condition is invariant under the allowable Poincar\'e-transformations..

First we want to formulate an algorithm to describe ideal, infinitesimal short
measurements playing the role of the reduction-postulate in the non-relativistic quantum
mechanics.

There should be $n$ measurements, which happen at the proper times $\tau_i$ at
the space-time points $z_i$, $i=1..n$. The $i$th measurement is represented by an
observable
\begin{eqnarray}
  A_i = \sum_j \lambda_{i,j} |\Phi_{i,j}><\Phi_{i,j}|
\end{eqnarray}
$\Phi_{i,j}$ being eigenvectors of $A_i$ with $1 = \sum_j |\Phi_{i,j}><\Phi_{i,j}|$ and
$<\Phi_{i,j}|\Phi_{i,k}> = \delta_{j,k}$.

We assume, that $\tau_i < \tau_j$ for $i<j$ and no ''later'' measurement take
place in the backward light-cone of a previous measurement (see above \refeq{eq_order}):

We can now formulate a relativistic reduction-postulate for ideal measurements:

\begin{itemize}
\item[(i)] The particle is prepared at a proper time $\tau_0$ with $\tau_0 <
  \tau_1$ at a space-time point $x_0$, the state of the quantum part should be $\Psi_{\tau_0}$
  with $<\Psi_{\tau_0}|\Psi_{\tau_0}> = 1$ and the classical state is
  $\omega_{\tau_0} = 0$. Set $i=1$.

\item[(ii)] The quantum and classical state change only in case of measurement, they have
no $\tau$-development if there is no measurement:
\begin{eqnarray}
  \Psi_\tau & = & \Psi_{\tau_{i-1}} \\
  \omega_\tau & = & \omega_{\tau_{i-1}}
\end{eqnarray}
for $\tau_{i-1} \le \tau \le \tau_{i}$. 

\item[(iii)] The $i$th measurement takes place at proper time $\tau_i$ at a
  space-time point $z_i$, we get
  the measurement result
  $\lambda_{i,j}$ with probability
\begin{eqnarray}
  p(\lambda_{i,j}) = \fabsq{<\Phi_{i,j}|\Psi_{i-1}>}
\end{eqnarray}
If $\lambda_{i,j}$ is the measurement result, the state of system after the
measurement is
\begin{eqnarray}
  \Psi_{\tau_i} & \longrightarrow & \Phi_{i,j} \\
  \omega_{\tau_i} & \longrightarrow & j
\end{eqnarray}

\item[(iv)] We set $i \to i+1$ and go to step (ii).
\end{itemize}

The probabilities generated by this algorithm are the same we get if we use the
standard-non-covariant reduction-postulate with the Dirac-equation and assume,
that space-like separated observable commute.

Now we formulate an algorithm for continuous relativistic measurements, indeed we
will propose in the following an algorithm to describe detections of an electron.

The electron should be prepared in a point $x_0=(x_0^0,\vec{x_0})$.

We consider $N$ detectors with trajectories $z_j (\tau)$, $j=1..N$. The
trajectories start at proper time $\tau = 0$ from the backward light cone of the 'preparing
event' $x_0$, $\fnormq {x_0 - z_j (0)} = 0, z_j^0(0) \le x_0^0$.
 
We allow detections, which happen in the past of the preparation time,
but we do not allow detections, if the detection space-time point lies in the
backward light-cone of the preparation event.

The coupling between the quantum and the classical system is given by operators
$G_j (\tau)$. We set $\Lambda(\tau) = \sum_j G_j (\tau)^+ G_j (\tau)$, $G_j
(\tau)^+$ being the adjoint operator.

An operator $G_j (\tau)$ is uniquely given by its operation on a projection of the
quantum state on a plane
$G_j (\tau) = U((y^0, \vec{y}), \vec{\alpha},\vec{\varphi}) \; g_j(\tau) \;
U^{-1}((y^0, \vec{y}), \vec{\alpha},\vec{\varphi})$ with $(y_0, \vec{y}),
\vec{\alpha}, \vec{\varphi}$ arbitrary.

We define now the following algorithm:

\begin{itemize}
\item[(i)] The particle is prepared in a point $x_0$, the state of the quantum
  part should be $\Psi_0$ with $<\Psi_0|\Psi_0> = 1$ and the classical state is
  $\omega = 0$, $\tau = 0$.

\item[(ii)] Choose uniformly a random number $r \in [0,1]$.

\item[(iii)] Propagate the quantum state forward in proper time by solving
\begin{eqnarray}
  \frac{\partial}{\partial \tau} \Psi_\tau = -\frac{1}{2} \Lambda(\tau)
  \Psi_\tau
\end{eqnarray}
until $\tau = \tau_1$, where $\tau_1$ is defined by
\begin{eqnarray}
  1 - <\Psi_{\tau_1}|\Psi_{\tau_1}> = \int_0^{\tau_1} d\tau
  <\Psi_\tau|\Lambda\Psi_\tau> = r
\end{eqnarray}
A detection happens at proper time $\tau = \tau_1$.

\item[(iv)] We choose the detector $k$, which detects the particle with
  probability
\begin{eqnarray}
  p_k = \frac{1}{N} <G_k(\tau_1) \Psi_{\tau_1} | G_k (\tau_1) \Psi_{\tau_1}>
\end{eqnarray}
with $N = \sum_j <G_j (\tau_1)\Psi_{\tau_1}| G_j(\tau_1)\Psi_{\tau_1}>$.

\item[(v)] Let $l$ be the detector, which detects the particle. The
  detection happens at the point $z_l (\tau_1)$. The detection induces then the
  following change of the states:

\begin{eqnarray}
  \Psi_{\tau_1} & \longrightarrow & \frac{G_l (\tau_1) \Psi_{\tau_1}}
  {\fsqrt{<G_l (\tau_1) \Psi_{\tau_1}|G_1 (\tau_1) \Psi_{\tau_1}>}} \\ \omega
  &\longrightarrow& l
\end{eqnarray}
\end{itemize}

The algorithm starts again perhaps with other detectors at position (ii).

The non-relativistic limit of this algorithm is the PDP-algorithm of the
EEQT. To prove this, we define
\begin{eqnarray}
 \Omega (\tau, \vec{x}) :=
 (U^{-1} ((c\tau + x_0^0, \vec{x_0}),\vec{0},\vec{0}) \Psi_\tau) (\vec{x})
 = \Psi_{\tau} (c\tau + x_0^0, \vec{x} + \vec{x_0})
\end{eqnarray}
We get
\begin{eqnarray}
  \lefteqn{\imath \hbar \frac{\partial}{\partial \tau}\Omega (\tau,\vec{x})}
 \nonumber \\
 & = &
  \imath \hbar c \left( \frac{\partial}{\partial x^0} \Psi_\tau \right)
  (c\tau+x_0,\vec{x}+\vec{x_0}) + \imath \hbar \frac{\partial \Psi_\tau}{\partial\tau}
  (c\tau+x_0,\vec{x}+\vec{x_0}) \nonumber \\
 & = &
  \imath \hbar c \left[ -\gamma^0 \gamma^k
  \frac{\partial \Psi_\tau}{\partial x^k} (c\tau+ x_0, \vec{x}+\vec{x_0}) 
  - \imath \frac{e}{c\hbar} \gamma^0 \gamma^\mu A_\mu - \imath
  \frac{mc}{\hbar}\gamma^0 \Psi_\tau (c\tau+ x_0, \vec{x}+\vec{x_0}) \right]
 \nonumber \\
 & &
  - \imath
  \frac{\hbar}{2} (U^{-1} \Lambda(\tau)\Psi_\tau) (\vec{x}) \nonumber \\
 & = &
  - \imath\hbar c \gamma^0\gamma^k \frac{\partial\Omega}{\partial x^k} + mc^2
  \gamma^0 \Omega + eA^0 \Omega - e \gamma^0 \vec{\gamma} \vec{A} \Omega
 \nonumber \\
 & &
  - \imath \frac{\hbar}{2} \left(\sum_j \underbrace{(U^{-1} G_j^+(\tau) U)}_{g_j^+(\tau)}
  \underbrace{(U^{-1} G_j (\tau) U)}_{g_j(\tau)} \right) \Omega 
\label{eq1}
\end{eqnarray}
with $U^{-1} = U^{-1} ((c\tau + x_0^0, \vec{x_0}),\vec{0},\vec{0})$ and $U = U
((c\tau + x_0^0, \vec{x_0}),\vec{0},\vec{0})$. We examine the non-relativistic limit
of \refeq{eq1} doing the assumption (compare to calculations of the limit using the
Dirac-equation in textbooks)
\begin{eqnarray}
\Omega (\tau, \vec{x}) = \fexp{-\imath \frac{mc^2}{\hbar} \tau} \left(
\begin{array}{c} \phi \\ \chi \end{array} \right)
\end{eqnarray}
Moreover we assume, that the detectors detect only particles
\begin{eqnarray}
g_j(\tau) = \left( \begin{array}{cc} \tilde{g_j}(\tau) & 0 \\ 0 & 0 \end{array}
\right)
\end{eqnarray}
In the non-relativistic limit we get the modified equation of the EEQT
\begin{eqnarray}
 \imath \hbar \frac{\partial}{\partial \tau}\phi
 & = & \left[ \frac{1}{2m} \sum_l \left( \frac{\hbar}{\imath} \frac{\partial}{\partial
 x^l} - \frac{e}{c} A^l \right)^2 - \frac{e\hbar}{2mc} \vec{\sigma} \vec{B}
 + e A^0 - \imath \frac{\hbar}{2} \sum_j \tilde{g_j}^+(\tau)
 \tilde{g_j}(\tau) \right] \phi
\end{eqnarray}
Noting, that $<\Psi_\tau | \Psi_\tau> \, = \int d\vec{x} \, \Omega^+ (\tau,\vec{x})
\Omega (\tau, \vec{x}) \stackrel{c\to\infty}{\longrightarrow} \int d\vec{x} \, \Phi^+
(\tau, \vec{x}) \Phi (\tau, \vec{x})$ and set $t := \tau$ we get the
PDP-algorithm of the EEQT as the non-relativistic limit of the above
relativistic algorithm.

%
%

\section {Application: Time of Arrival}

One application of the above algorithm is the detection of a particle by one
detector, which is at rest. We look at the problem in $1+1$ dimension.

We introduce the operator $U^{-1}(y^0, y^1)$
\begin{eqnarray}
  (U^{-1} (y^0, y^1) \Psi) (x) := \Psi (y^0, y^1 + x)
\end{eqnarray}
and the inverse operator $U(y^0, y^1)$. The operators are unitary.

The particle is prepared at $(c t_0 \equiv 0, x_0 \equiv -1\Ang)$ and the
initial state of the particle should be $\Psi_0 = U(c t_0, x_0)\psi_0$ with
\begin{eqnarray}
  \psi_0 (x) = \frac{1}{(2\pi)^{1/4}\fsqrt{\eta}} \cdot \fexp{-\frac{x^2}{4\eta^2} + \imath
    \frac{p_0}{\hbar}x}\cdot \left(\begin{array}{c}1\\0\\0\\0\end{array}\right)
\end{eqnarray}
and $\eta = 0.1 \Ang$. We get

\begin{eqnarray}
\lefteqn{\Psi_0 (ct,x) =} \nonumber \\
& & \frac{\fsqrt{2\eta}}{(2\pi)^{1/4}} \int dp \,
\frac{E+mc^2}{2E} \fexp{-\eta^2 \frac{(p-p_0)^2}{\hbar^2} - \imath
  \frac{p}{\hbar}x_0} \left(\begin{array}{c}1\\0\\0\\\frac{pc}{E+mc^2}\end{array}\right)
\cdot \fexp{\imath \frac{p}{\hbar} x - \imath \frac{E}{\hbar}t} \nonumber\\
& & + \frac{\fsqrt{2\eta}}{(2\pi)^{1/4}} \int dp \,
\frac{pc}{2E} \fexp{-\eta^2 \frac{(p+p_0)^2}{\hbar^2} + \imath
  \frac{p}{\hbar}x_0} \left(\begin{array}{c}\frac{pc}{E+mc^2}\\0\\0\\1\end{array}\right)
\cdot \fexp{-\imath \frac{p}{\hbar} x + \imath \frac{E}{\hbar}t}
\end{eqnarray}

\begin{figure}
  \begin{center}
    \caption{}
    {Initial state for particle momentum $p_0=0.75$} \leavevmode
    \includegraphics [width=0.9\linewidth]{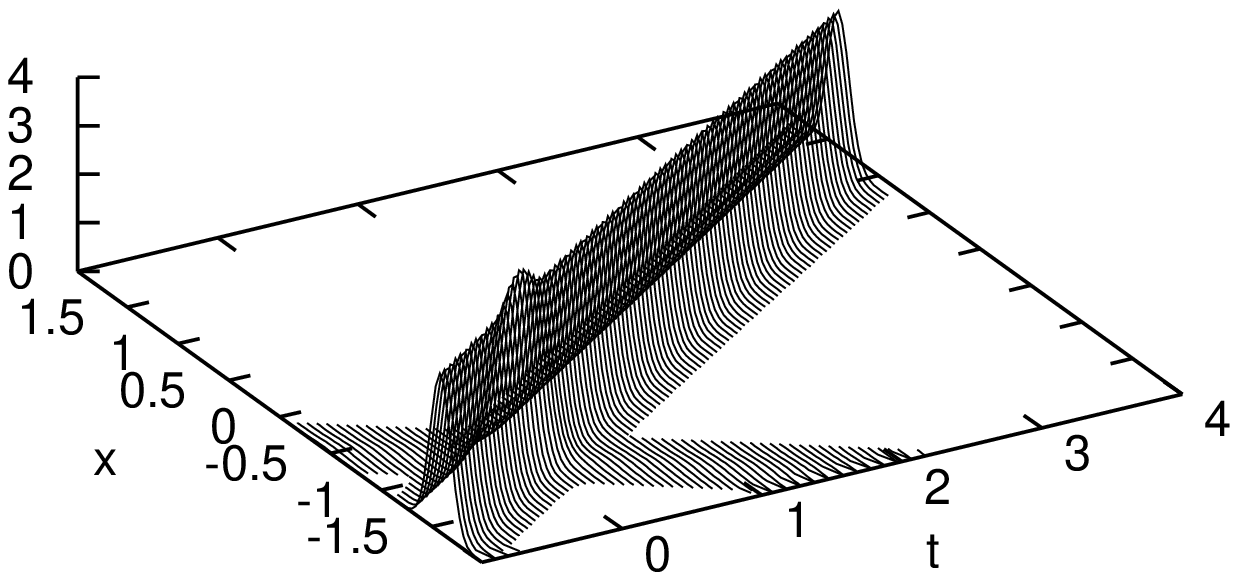}

    {Square of the first component of the wave function (particle part)}
    \leavevmode
    \includegraphics [width=0.9\linewidth]{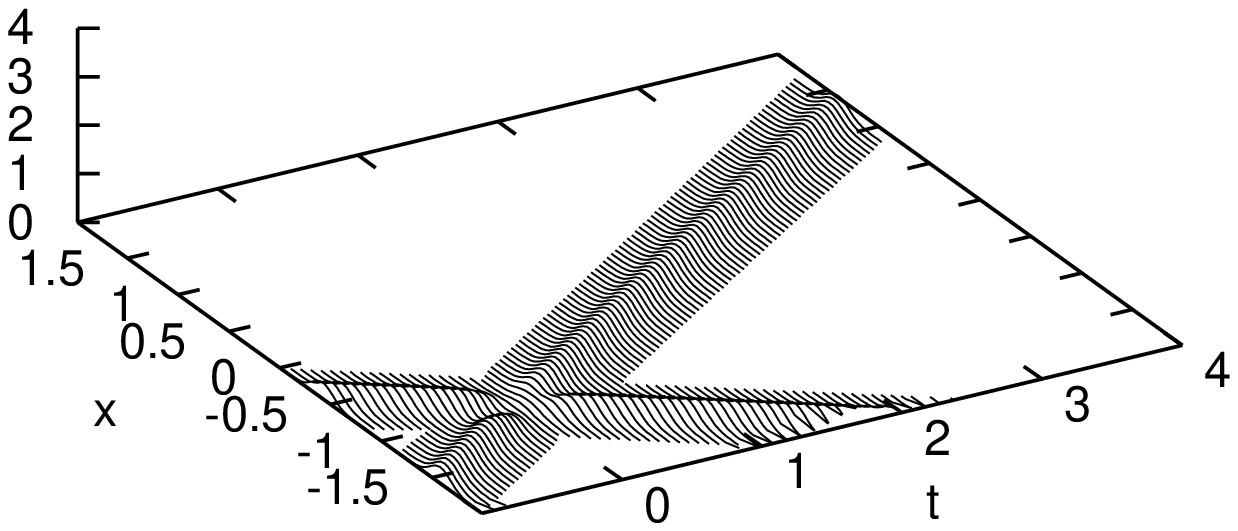}

    {Square of the forth component of the wave function (anti-particle part)}
    \label{fig_1}
  \end{center}
\end{figure}

For illustration, Fig. \ref{fig_1}~(a) shows the square of the first component
(particle) and Fig. \ref{fig_1}~(b) the square of the forth component
(anti-particle) of the initial state $\Psi_0 (ct,x)$ for the momentum $p_0 =
0.75 mc$.

The detector is put at $x_D = 0\Ang$, its trajectory is $z(\tau) = (c \tau+x_0,
x_D)$.

The coupling operator of the detector is given by
\begin{eqnarray}
  G (\tau) = U(z (\tau))g(x)U^{-1}(z(\tau))
\end{eqnarray}
with $g(x)$ a function characterizing the sensitivity of the detector.

The detector should detect only particles and not anti-particles.

Using our algorithm, the total probability, that the detector detects the
electron is given by
\begin{eqnarray}
  P_{\infty} = \int_0^\infty d\tau \, <\Psi_\tau | \Lambda \Psi_\tau >
\end{eqnarray}
The probability density for a ''proper time of arrival'' at the detector is
given by
\begin{eqnarray}
  P(\tau) = \frac{1}{P_\infty} <\Psi_\tau | \Lambda \Psi_\tau >
\end{eqnarray}

With this probability density we can calculate the probability density for the
time of arrival and the expected value in each system, for example in a system
in which the detector is at rest:
\begin{eqnarray}
  p(t) & = & P(t- \frac{x_0}{c}) \\ T & = & \int dt \, t \cdot p(t) = \int d\tau
  \, \left( \tau + \frac{x_0}{c} \right) P(\tau) = \int d\tau \, \tau P(\tau) +
  \frac{x_0}{c}
\end{eqnarray}

Now we want to examine how the same situation looks like in an reference frame $\tilde{K}$,
which moves with velocity $v$ with respect to the system $K$ in which the detector is at
rest. The transformation has the following form:
\begin{eqnarray}
  \tilde{x} = \frac{1}{\fsqrt{1-\frac{v^2}{c^2}}} \left(
\begin{array}{cc} 1&\frac{v}{c}\\\frac{v}{c}&1\end{array}
\right) x
\end{eqnarray}

The normalized probability density for the time of arrival is given by
\begin{eqnarray}
  \tilde{p}(\tilde{t}) = \fsqrt{1-\frac{v^2}{c^2}} p \left(
  \fsqrt{1-\frac{v^2}{c^2}} \tilde{t} \right) = \fsqrt{1-\frac{v^2}{c^2}} P
  \left( \fsqrt{1-\frac{v^2}{c^2}} \tilde{t} - \frac{x_0}{c} \right)
\end{eqnarray}

The expected value is given by
\begin{eqnarray}
  \tilde{T} & = & \int d\tilde{t} \, \tilde{t} \, \tilde{p}(\tilde{t}) =
  \fsqrt{1-\frac{v^2}{c^2}} \int d\tilde{t} \, \tilde{t} \,
  p\left(\fsqrt{1-\frac{v^2}{c^2}} \tilde{t}\right) \nonumber \\ & = &
  \fsqrt{1-\frac{v^2}{c^2}} \int dt \, \frac{1}{\fsqrt{1-\frac{v^2}{c^2}}} \,
  \frac{t}{\fsqrt{1-\frac{v^2}{c^2}}} \, p(t) = \frac{1}{\fsqrt{1-
      \frac{v^2}{c^2}}} T
\label{conn_diff_IS}
\end{eqnarray}

For computation of $P(\tau)$ we define
\begin{eqnarray}
  \Omega (\tau, x) := (U^{-1}(z(\tau)) \Psi_\tau) (x) = \Psi_\tau (c \tau +
  x_0, x + x_D)
\end{eqnarray} 
and we note, that $<\Psi_\tau|\Psi_\tau> = \int dx \fabsq{\Omega(\tau,x)}$ or
$<\Psi_\tau|\Lambda \Psi_\tau> = \int dx \Omega^+(\tau,x) g^+(x) g(x)\Omega(\tau,x)$.

We must now solve
\begin{eqnarray}
 \imath \hbar \frac{\partial}{\partial \tau}\Omega (\tau,x)
 & = &
  - \imath\hbar c \gamma^0\gamma^1
  \frac{\partial\Omega}{\partial x} + mc^2 \gamma^0 \Omega - \imath
  \frac{\hbar}{2} g^+(x) g(x) \Omega
\label {dgl_omega}
\end{eqnarray}
with the initial condition $\Omega (0,x)=\Psi_0 (x_0, x+x_D) = \Psi_0 (x_0,x)$.

The probability density for the ''proper time of arrival'' is now
\begin{eqnarray}
P(\tau) = \frac{1}{\int_0^\infty d\tau \, \int dx \, \Omega^+(\tau,x) g^+(x)
  g(x)\Omega(\tau,x)} \cdot \int dx \, \Omega^+(\tau,x) g^+(x)
  g(x)\Omega(\tau,x)
\end{eqnarray}

%
%

\subsection{Time of arrival with wide detectors}

The detector should be characterized by the sensitivity function
\begin{eqnarray}
  g (x) = \fsqrt{\frac{2W}{\hbar}} \left( \begin{array}{cccc}
  1&0&0&0\\0&1&0&0\\0&0&0&0\\0&0&0&0\end{array}\right) \cdot
  \left\{\begin{array}{c@{\quad:\quad}l} 0 & x < -\frac{\Delta x_D}{2} \\ 
    \fexp{-\frac{\left(-\frac{\Delta x_D}{2} + \epsilon - x\right)^2}{\epsilon^2
        - \left(\frac{-\Delta x_D}{2} + \epsilon - x\right)^2}} & -\frac{\Delta
      x_D}{2} \le x < -\frac{\Delta x_D}{2} + \epsilon \\ 1 & -\frac{\Delta
      x_D}{2} + \epsilon \le x < \frac{\Delta x_D}{2} - \epsilon \\ 
    \fexp{-\frac{\left(x-\frac{\Delta x_D}{2} + \epsilon\right)^2}{\epsilon^2 -
        \left(x - \frac{-\Delta x_D}{2} + \epsilon\right)^2}} & \frac{\Delta
        x_D}{2} - \epsilon \le x < \frac{\Delta x_D}{2} \\ 0 & \frac{\Delta
        x_D}{2} \le x
\end{array} \right.
\end{eqnarray}
with $\epsilon = 0.002 \Ang$.
 
The \refeq{dgl_omega} with the initial condition $\Omega (0, x) = 
 \Psi_0 (x_0, x)$ is solved numerically. The time development of $\Omega$ is approximated by
\begin{eqnarray*}
  \lefteqn{\Omega(\tau+\Delta \tau)} \\ & = & \fexp{-\Delta\tau
    \frac{\imath}{\hbar}\left(-\imath\hbar c \gamma^0 \gamma^1
    \frac{\partial}{\partial x^1} + mc^2 \gamma^0 \right) - \Delta\tau g^+(x)
    g(x)} \Omega (\tau) \\ & \approx & \fexp{-\frac{\Delta\tau}{2}g^+(x) g(x)}
  \fexp{-\Delta\tau \frac{\imath}{\hbar}\left(-\imath\hbar c \gamma^0\gamma^1
    \frac{\partial}{\partial x^1} + mc^2 \gamma^0 \right)} \\ & &
  \fexp{-\frac{\Delta\tau}{2}g^+(x) g(x)}\Omega (\tau)
\end{eqnarray*}

We now discretize the proper time and the space. The first and the last operator can
then be computed directly, the second operator is discretized by using the method of
Wessels, Caspers and Wiegel \cite{wessels.1999}.

The boundary conditions are walls at $x=-6 \Ang$ and at $x=4
\Ang$, the time and space steps depend on the particle momentum: $p_0 = 0.25-0.75
(\Delta\tau = \Delta x = 0.001), p_0 = 1.0 (\Delta\tau = \Delta x = 0.00075),
p_0 = 1.25 (\Delta\tau = \Delta x = 0.00075), p_0 = 1.5 (\Delta\tau = \Delta x =
0.0005), p_0 = 1.75 (\Delta\tau = \Delta x = 0.00043), p_0 = 2.0 (\Delta\tau =
\Delta x = 0.000375)$.

\begin{figure}
  \begin{center}
    \leavevmode
    \includegraphics [width=0.9\linewidth]{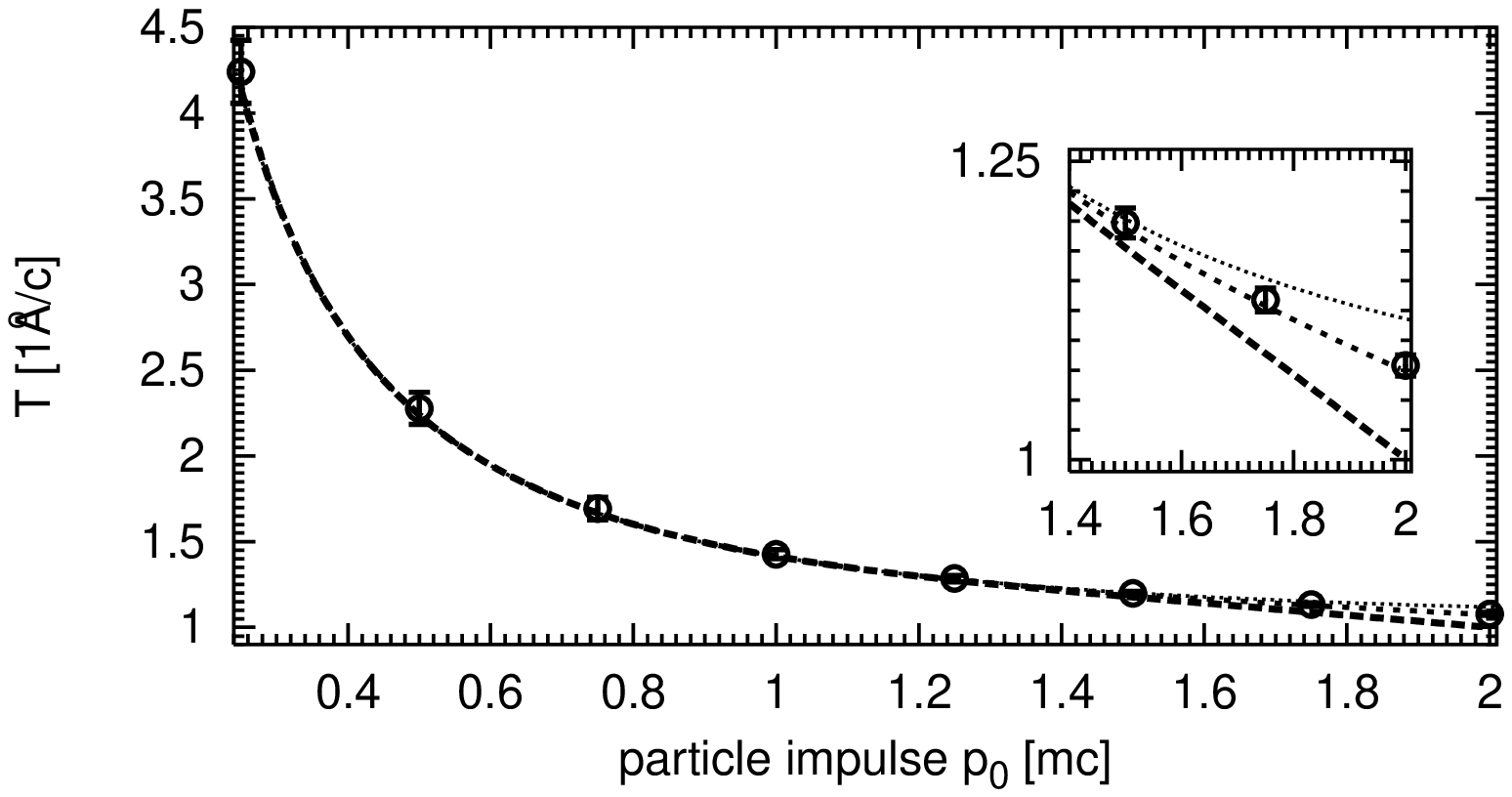}

    \caption{}
    {Mean time of arrival versus particle momentum $p_0$, relativistic
      simulation with wide detector (circles with errorbars): detector height
      $W=1\times 10^{-5}\,mc^2$, detector width $\Delta x_D = 0.01 \Ang$, other
      parameters see text, point-like detector: $\kappa \to 0$ (big dotted line),
      $\kappa=1.0 c/1\Ang$ (dashed line), relativistic mechanics $1\Ang \cdot \fsqrt{1 +
      \frac{1}{p_0^2}}$ (small dotted line);
    the figure inside is a zoom of the right lower area of the figure outside}
  
    \label{fig_2}
  \end{center}
\end{figure}

Fig.\ref{fig_2} shows the resulting expected values of the time of arrival for
different momentums $p_0$ in the system in which the detector is at rest.  Moreover
the times calculated by the relativistic mechanics of point-particle are shown.

The error is approximated by
\begin{eqnarray}
  error(T) = \fabs{\frac{1}{\lambda-1}} \fabs{T(\lambda \cdot \Delta\tau) -
    T(\Delta\tau)}
\end{eqnarray}
with $\lambda=1.5$.

We find a good coincidence between the simulated results and those of the
mechanics of point-particles. Only for very high momentums the expected times of
the simulations are a bit smaller than those of the point-mechanics, because
there is a probability for negative times of arrival in the simulations (see below).

The expected values in different reference frames are connected by
\refeq{conn_diff_IS}. The time of arrival $\tilde{t}_{RM}$ of the relativistic
mechanic is connected
to the result for a detector at rest $t_{RM}$ of the mechanic by the same
factor:
\begin{eqnarray}
  \tilde{t}_{RM} = \frac{1}{\fsqrt{1-\frac{v^2}{c^2}}} t_{RM}
\end{eqnarray}
  
So we also have a good coincidence between the simulated results and those
deduced from the relativistic mechanics in all reference frames.

\begin{figure}
  \begin{center}
    \leavevmode
    \includegraphics [width=0.9\linewidth]{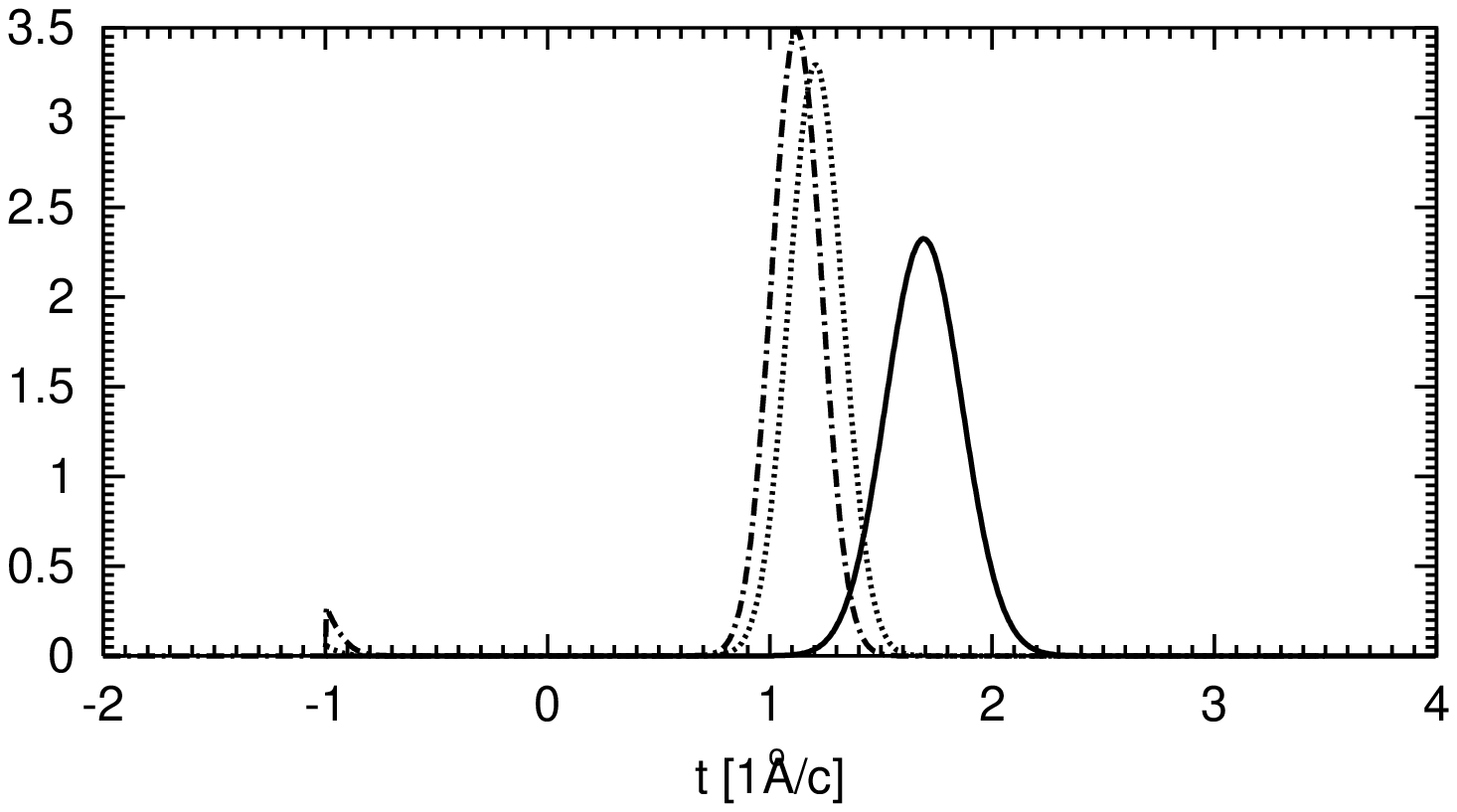}

    \caption{}
    {Probability density for the time of arrival for different particle momentum
      $p_0$, detector height $W=1\times 10^{-5}\, mc^2$, detector width $\Delta
      x_D = 0.01 \Ang$, $p0 = 0.75 mc$ (solid line), $p0 = 1.5 mc$ (dotted line),
      $p0 = 2.0 mc$ (dashed-dotted line)}
    \label{fig_4}
  \end{center}
\end{figure}

In Fig.\ref{fig_4} probability densities in the system in which the detector is
at rest are shown. The
probability for negative times of arrival is zero for small momentums, but for
momentums with are greater than one, we find a small probability for negative
times.

\begin{figure}
  \begin{center}
    \leavevmode
    \includegraphics [width=0.9\linewidth]{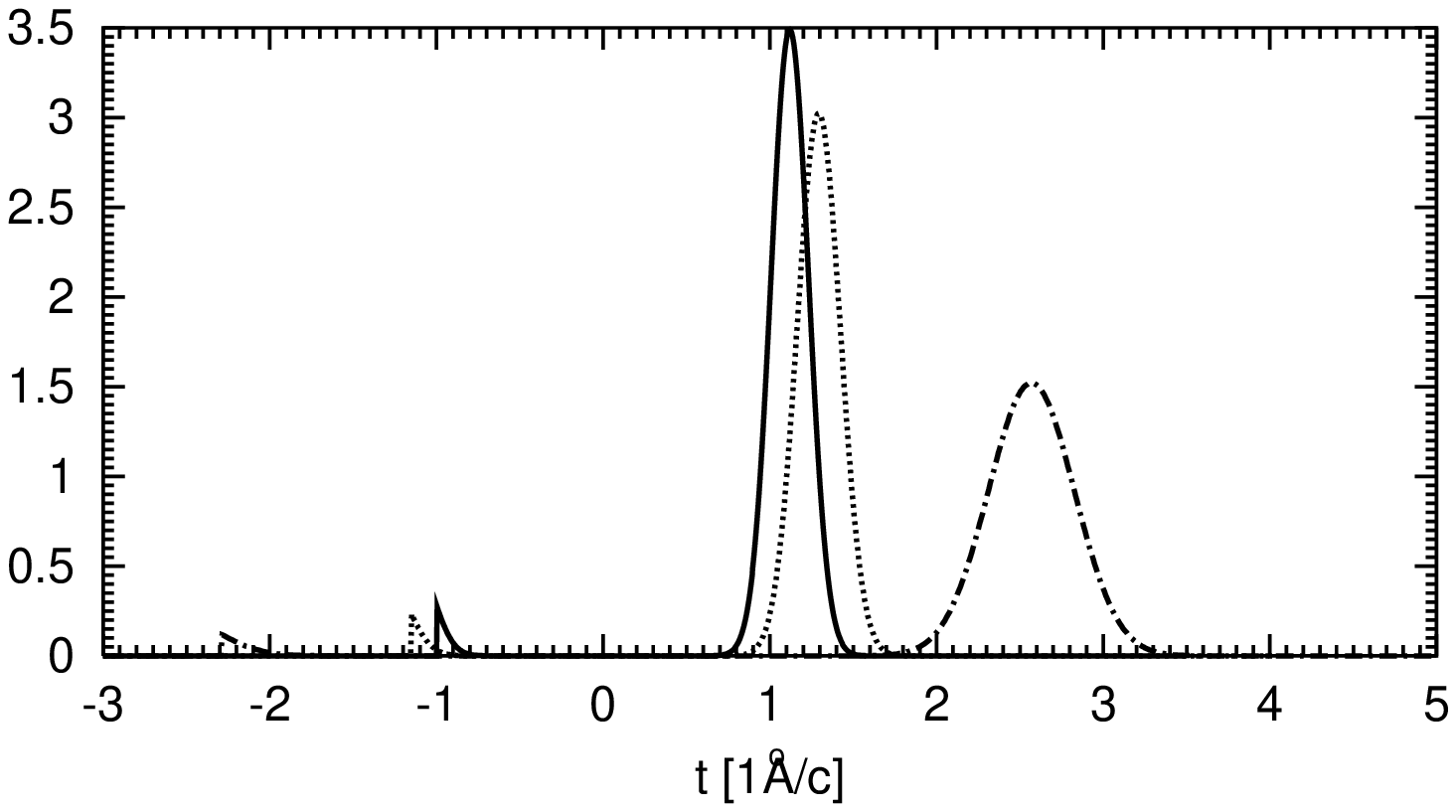}

    \caption{}
    {Probability density for the time of arrival in different reference frames,
      particle momentum $p_0=2.0\,mc$, detector height $W=1\times 10^{-5}\,mc^2$,
      detector width $\Delta x_D=0.01 \Ang$, $v=0.0 c$ (solid line), $v=0.5 c$
      (dotted line), $v=0.9 c$ (dashed-dotted line)}
    \label{fig_10}
  \end{center}
\end{figure}

Fig.\ref{fig_10} show the probability density for different system velocities
$v$ with fixed particle energy $p_0 = 2.0 mc$.

Another question is, how the expected times depend on parameters of the
detector. The detector width should be fixed at $\Delta x_D =
0.01 \Ang$ and the particle momentum is $p_0 = 0.75 mc$.  We find, that the
expected values
for this parameters are nearly independent of the detector height $W$ over a
wide range, only the total
detection probability depends on the detector height $W$.

The normalized probability densities are also independent of the detector
height.

Moreover we examine the dependence, if the width of the detector is changed with
fixed detector height $W=1\times 10^{-5}mc^2$ and particle momentum $p_0 = 0.75mc$. The
expected values also show nearly dependence, the total detection
probability shows a dependence on the detector width. The form of the
probability densities
do not change, they only becomes wider.
\newpage

%
%

\subsection {Time of arrival with point-like detectors}

In this section, we will examine the limit of a point-like detector. The
detector sensitivity should be characterized by

\begin{eqnarray}
g^+ (x) g(x) = \kappa \cdot \left( \begin{array}{cccc}
  \delta(x)&0&0&0\\0&\delta(x)&0&0\\0&0&0&0\\0&0&0&0\end{array}\right)
\label{det_delta}
\end{eqnarray}

A non-relativistic particle-detector modelized by a
$\delta$-function can be found in \cite{blanchard.1996a}.
 
By integration of \refeq{dgl_omega} with the detector function
\refeq{det_delta}: $\int_{-\epsilon}^{\epsilon} dx$ and do $\epsilon \to 0$, we
get the boundary conditions at $x=0$:
\begin{eqnarray}
\Omega_1 (\tau, 0^+) - \Omega_1 (\tau, 0^-)  & =  & 0 \\
\Omega_2 (\tau, 0^+) - \Omega_2 (\tau, 0^-)  & =  & 0 \\
\Omega_3 (\tau, 0^+) - \Omega_3 (\tau, 0^-)  & =  & - \frac{\kappa}{2c} \Omega_2
(\tau, 0) \\
\Omega_4 (\tau, 0^+) - \Omega_3 (\tau, 0^-)  & =  & - \frac{\kappa}{2c} \Omega_1
(\tau, 0)
\end{eqnarray}

A solution of \refeq{dgl_omega} with the detector function
\refeq{det_delta} with these boundary conditions is now
\begin{eqnarray}
\Omega (\tau, x) = \left\{
\begin{array}{lcl} \Omega_{IN} (\tau,x) +
\Omega_{REF} (\tau,x) & : & x < 0 \\
\Omega_{TRA} (\tau,x) & : & x > 0
\end{array}
\right.
\end{eqnarray}
with
\begin{eqnarray*}
%
%
\lefteqn{\Omega_{IN} (\tau,x) =} \nonumber \\
& & \int dp \, \left[ 
A_+ (p) \left(\begin{array}{c}1\\0\\0\\\frac{pc}{E+mc^2}\end{array}\right)
+ A_- (p) \left(\begin{array}{c}0\\1\\\frac{pc}{E+mc^2}\\0\end{array}\right)
\right]
\cdot \fexp{\imath \frac{p}{\hbar}x - \imath \frac{E}{\hbar} \tau} \nonumber \\
& & + \int dp \, \left[ 
B_+ (p) \left(\begin{array}{c}\frac{pc}{E+mc^2}\\0\\0\\1 \end{array}\right)
+ B_- (p) \left(\begin{array}{c}0\\\frac{pc}{E+mc^2}\\1\\0\end{array}\right)
\right]
\cdot \fexp{-\imath \frac{p}{\hbar}x + \imath \frac{E}{\hbar} \tau} \\
%
%
\lefteqn{\Omega_{REF} (\tau,x) =} \nonumber \\
& & \int dp \, 
\frac{(\kappa/2c)(E+mc^2)}{2pc-(\kappa/2c)(E+mc^2)} 
\left[ 
A_+ (-p) \left(\begin{array}{c}1\\0\\0\\\frac{pc}{E+mc^2}\end{array}\right)
+ A_- (-p) \left(\begin{array}{c}0\\1\\\frac{pc}{E+mc^2}\\0\end{array}\right)
\right]
\cdot \fexp{\imath \frac{p}{\hbar}x - \imath \frac{E}{\hbar} \tau} \nonumber \\
& & + \int dp \, 
\frac{pc (\kappa/2c)}{pc (\kappa/2c) - 2(E+mc^2)}
\left[ 
B_+ (-p) \left(\begin{array}{c}\frac{pc}{E+mc^2}\\0\\0\\1 \end{array}\right)
+ B_- (-p) \left(\begin{array}{c}0\\\frac{pc}{E+mc^2}\\1\\0\end{array}\right)
\right]
\cdot \fexp{-\imath \frac{p}{\hbar}x + \imath \frac{E}{\hbar} \tau} \\
%
%
\lefteqn{\Omega_{TRA} (\tau,x) =} \nonumber \\
& & \int dp \, 
\frac{2pc}{2pc+(\kappa/2c)(E+mc^2)} 
\left[ 
A_+ (p) \left(\begin{array}{c}1\\0\\0\\\frac{pc}{E+mc^2}\end{array}\right)
+ A_- (p) \left(\begin{array}{c}0\\1\\\frac{pc}{E+mc^2}\\0\end{array}\right)
\right]
\cdot \fexp{\imath \frac{p}{\hbar}x - \imath \frac{E}{\hbar} \tau} \nonumber \\
& & + \int dp \, 
\frac{2(E+mc^2)}{pc (\kappa/2c) + 2(E+mc^2)}
\left[ 
B_+ (p) \left(\begin{array}{c}\frac{pc}{E+mc^2}\\0\\0\\1 \end{array}\right)
+ B_- (p) \left(\begin{array}{c}0\\\frac{pc}{E+mc^2}\\1\\0\end{array}\right)
\right]
\cdot \fexp{-\imath \frac{p}{\hbar}x + \imath \frac{E}{\hbar} \tau}
\end{eqnarray*}

At $\tau = 0$ the reflection part is approximated to be small, so the initial
value is $\Omega_{IN} (0,x) = N \cdot \Psi_0 (x_0, x)$ with $N$ a normalization
factor. So we get

\begin{eqnarray}
A_+(p) & = & N \cdot \frac{2\eta}{(2\pi)^{1/4}} \frac{E+mc^2}{2E} \fexp{-\eta^2
  \frac{(p-p_0)^2}{\hbar^2} - \imath \frac{p}{\hbar} x_0 - \imath
  \frac{Ex_0}{\hbar c}} \\
A_-(p) & = & 0 \\
B_+(p) & = & N \cdot \frac{2\eta}{(2\pi)^{1/4}} \frac{pc}{2E} \fexp{-\eta^2
  \frac{(p+p_0)^2}{\hbar^2} + \imath \frac{p}{\hbar} x_0 + \imath
  \frac{Ex_0}{\hbar c}} \\ 
B_-(p) & = & 0
\end{eqnarray}

The probability density of the ''proper time of arrival'' is now easy calculated:

\begin{eqnarray}
P(\tau) = \frac{1}{\int_0^\infty d\tau \, \fabsq{\Omega_1(\tau,0)}}
\fabsq{\Omega_1(\tau,0)} = \frac{1}{\int_0^\infty d\tau \, \fabsq{\Omega_{TRA,1}(\tau,0)}}
\fabsq{\Omega_{TRA,1}(\tau,0)}
\end{eqnarray}

Fig. \ref{fig_2} shows also the expected values for the parameter $\kappa=1.0 c/1\Ang$
and the limit $\kappa \to 0$. The probability density $P(\tau)$ remains finite in
the limit $\kappa \to 0$ and the wave function becomes the undisturbed
Gauss-function. The results of $\kappa \to 0$ equals nearly those with wide detectors.

\begin{figure}
  \begin{center}
    \leavevmode
    \includegraphics [width=0.9\linewidth]{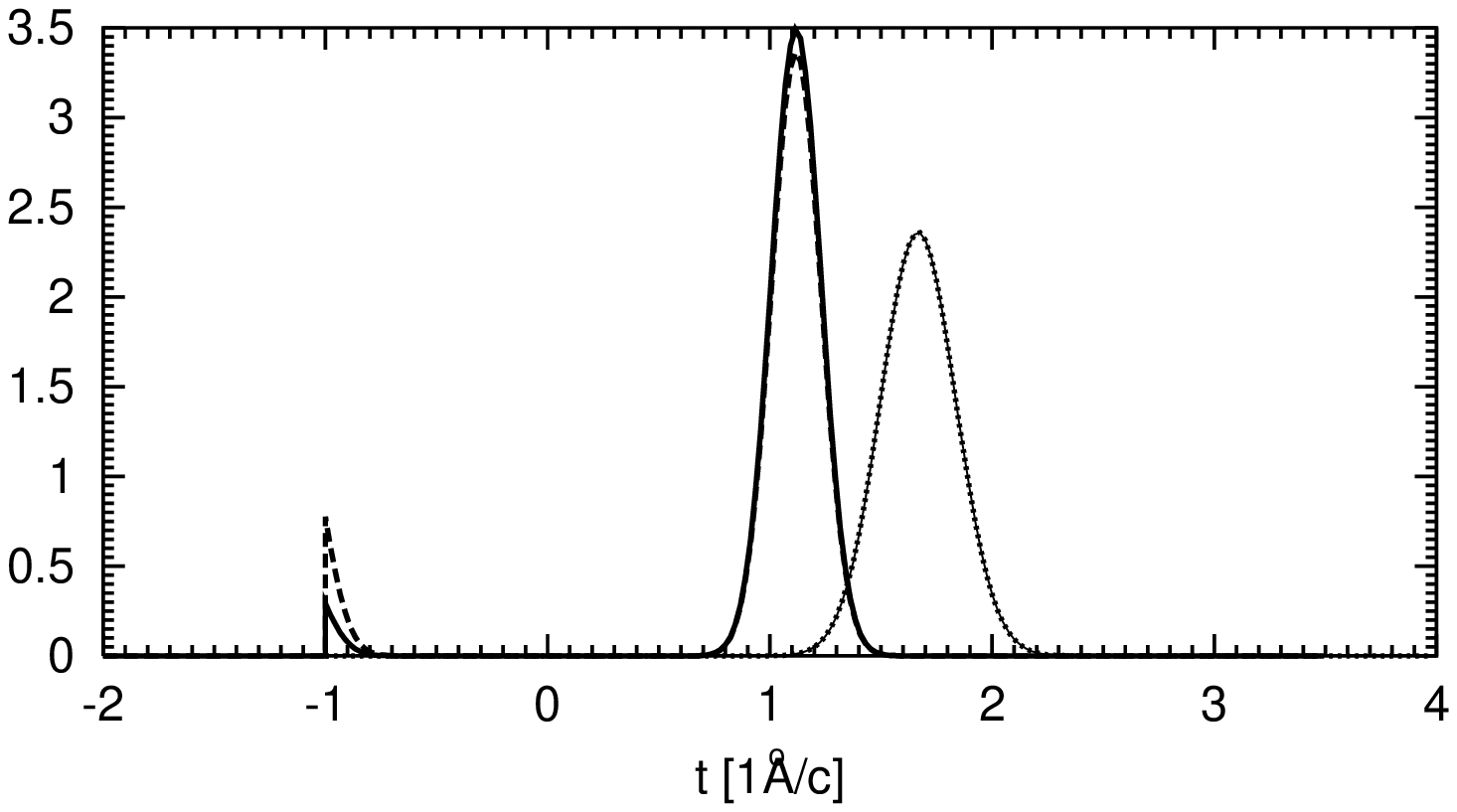}

    \caption{}
    {Probability density for the time of arrival for different particle momentum
      $p_0$ with point-like detector, $p0 = 0.75 mc, \kappa\to 0$ (small
      solid line), $p0 = 0.75 mc, \kappa=1.0 c/1\Ang$ (dotted line), $p0 = 2.0 mc, \kappa
      \to 0$ (big solid line),
      $p0 = 2.0 mc, \kappa=1.0 c/1\Ang$ (dashed line)}
    \label{fig_5}
  \end{center}
\end{figure}

The expected times are nearly independent of $\kappa$, only for high momentums the
time of arrival decreases with increasing $\kappa$. The
reason is an increasing of the probability of negative times of arrival with
increasing $\kappa$ (see Fig. \ref{fig_5}).

\section{Conclusion}

Summarizing we introduce a relativistic algorithm to describe ideal,
infinitesimal short measurements playing the role of the reduction-postulate in
the non-relativistic quantum mechanics.

Assuming that space-like separated observables commute, the probabilities
generated by this algorithm are the same we get if we use the
standard-non-covariant reduction-postulate with the Dirac-equation.

Moreover we introduce a relativistic algorithm for
continuous measurements, its
non-relativistic limit is the algorithm of the Event-Enhanced Quantum Theory.

We discuss an application of it, the time of arrival of an
electron at a detector.

First we examine numerically the case, if the detector is characterized by a
wide, finite high sensitivity function. In doing so we find a
good coincidence between the expected values of the time of arrival
which results by the algorithm and the results of the relativistic
point-mechanics.
For very high momentum we find a small probability for negative time of arrival,
so the expected values of the algorithm in these cases are a bit smaller than
the results expected by the relativistic point-mechanics.

The results do not depend sensitively on the detector parameters over a
wide range.

Second we examine the limit, if the detector sensitivity is characterized by a point-like,
infinitesimal high function $\sim \kappa \delta(x)$.
For weak detectors ($\kappa \to 0$) the same results as in the case of a wide
detector are found. For high momentums and $\kappa > 0$, the expected time
decreases with increasing detector ''height'' $\kappa$.
\\[0cm]

\noindent{\bf\large Acknowledgments}

I would like to thank Ph. Blanchard for many helpful discussions.
\newpage

%
%

%
%

\pagestyle{empty}

\end{document}